\documentclass[prb, superscriptaddress, showpacs, floatfix, twocolumn]{revtex4}
\usepackage{amsmath}
\usepackage{bbm}
\usepackage{graphicx}
\usepackage{float}
\usepackage{epsfig}
\usepackage{epsf}
\usepackage{bbold}
\usepackage{hyperref}

\usepackage{soul}
\usepackage{color}

\begin{document}
\title{Designing  ferromagnetism in vanadium-oxide based superlattices}
\author{Hung T. Dang}
\author{Andrew J. Millis}
\affiliation{Department of Physics, Columbia University, 538 West 120th Street, New York, New York 10027, USA}
\date{\today}

\begin{abstract}
Motivated by recent reports (Phys. Rev. B\textbf{80}, 241102) of room-temperature ferromagnetism in vanadium-oxide based superlattices, a single-site dynamical mean field study  of the  dependence of the paramagnetic-ferromagnetic phase boundary on superlattice geometry was performed. An examination of variants of the experimentally determined crystal structure indicate that  ferromagnetism  is found only in a small and probably inaccessible region of the phase diagram. Design criteria for increasing the range over which ferromagnetism might exist are proposed.
\end{abstract}

\maketitle

\section{Introduction\label{sec:intro}}
``Materials by design'', the ability to design and create a material with specified correlated electron properties, is a long-standing goal of condensed matter physics. Superlattices, in which one or more component is a transition metal oxide with a partially filled $d$-shell, are of great current interest in this regard because they offer the possibility of enhancing and controlling  the correlated electron phenomena known \cite{RevModPhys.70.1039} to occur in bulk materials as well as the possibility of creating electronic phases not observed in bulk.\cite{Millis11} Following the pioneering work of Ohtomo and Hwang,\cite{Ohtomo_Hwang} heterostructures and heterointerfaces of transition metal oxides have been studied extensively. Experimental findings include metal-insulator transitions,\cite{PhysRevLett.104.147601} superconductivity, \cite{Reyren07} magnetism \cite{PhysRevB.80.241102,Ariando11} and coexistence of ferromagnetic and superconducting phases.\cite{Bert11:_direct,Li11:_coexis}

\begin{figure}
 \includegraphics[width=\columnwidth]{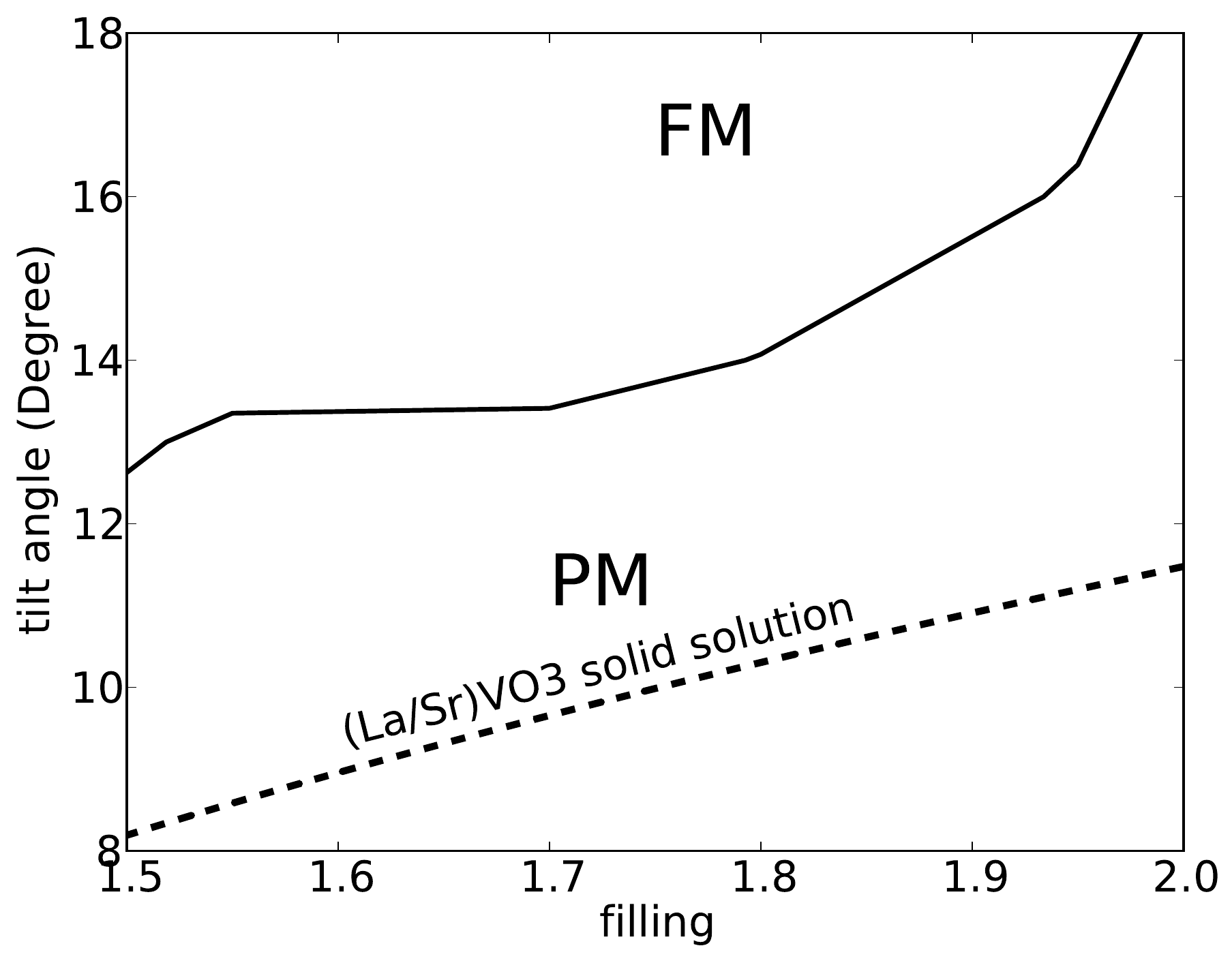}
 \caption{Ferromagnetic-paramagnetic phase diagram for La/SrVO$_3$ solid solution in plane of carrier concentration (changed by Sr concentration) and tilt angle in $Pnma$ structure but with all three Glazer's angles nearly equal. Dashed line indicates relation between carrier concentration and rotation amplitude in physically occurring bulk solid solution. From Ref.~\onlinecite{PhysRevB.87.155127}.}
\label{fig:bulkpd}
\end{figure}

In this paper we consider the possibility that appropriately designed superlattices might exhibit ferromagnetism. Our work is partly motivated by a recent report\cite{PhysRevB.80.241102} of room-temperature ferromagnetism in superlattices composed of some number $m$ of  layers of LaVO$_3$ (LVO) separated by one layer of  SrVO$_3$ (SVO), even though ferromagnetism is not found at any $x$ in the bulk solid solution La$_{1-x}$Sr$_x$VO$_3$. Our study is based on a previous analysis\cite{PhysRevB.87.155127} of the possibility of obtaining  ferromagnetism in variants of the crystal structure of  bulk solid solutions of the form La$_{1-x}$Sr$_x$VO$_3$. A key result of the previous work was that ferromagnetism is favored by a combination of large octahedral rotations and large doping away from the Mott insulating LaVO$_3$ composition. A schematic phase diagram is shown in Fig.~\ref{fig:bulkpd}. However, as indicated by the dashed line in the figure,  in the physical bulk solid solution, doping away from the Mott insulating concentration reduces the amplitude of the octahedral rotations so that the physical materials remain far from the magnetic phase boundary. The  motivating  idea of this paper is that in the superlattice geometry, octahedral rotation amplitude may be decoupled from carrier concentration. The rotations can be controlled by choice of substrate while the  carrier concentration can be controlled by choice of chemical composition and may vary from layer to layer of a superlattice. In effect, an appropriately designed superlattice could enable the exploration of  different paths in Fig.~\ref{fig:bulkpd}.

In this study, we combine single-site dynamical mean field approximation\cite{Georges96} with realistic band structure calculations including the effects of the octahedral rotations to determine the ferromagnetic-paramagnetic phase diagram in superlattices with the crystal structures believed relevant\cite{PhysRevB.83.125403,PhysRevB.85.184101} to the experiments of Ref. \onlinecite{PhysRevB.80.241102}. Unfortunately we find that the experimentally determined crystal structure is in fact less favorable to ferromagnetism than the one found in the bulk solid solution, but we indicate structures that may be more favorable. 

The paper has following structure. The model and methods  are described in Sec.~\ref{sec:model}. Sec.~\ref{sec:cubicsuperlattice} establishes the methods via a detailed analysis of the phase diagram of superlattices with no rotations or tilts. In Sec.~\ref{sec:tiltedsuperlattice} we present the magnetic properties of superlattices with octahedral rotations similar to those observed experimentally. Section~\ref{sec:conclusions} is a summary and conclusion.

\section{Model and methods\label{sec:model}}
\subsection{Overview}
This paper builds on a previous study of the  magnetic phase diagram of bulk vanadates.\cite{PhysRevB.87.155127} The new features relevant for the superlattices studied here  are (i) the change in geometrical structure, including the differences from the bulk solid solution in the  pattern of octahedral tilts and rotations and (ii) the variation of electronic density arising from superlattice structure.  In the rest of this section we briefly summarize the basic theoretical methodology (referring the reader to Ref.~\onlinecite{PhysRevB.87.155127} for details), define the crystal structures more precisely, explain the consequences for the electronic structure and explain how the variation of density appears in the formalism. 

\subsection{Geometrical structure}

We study superlattices  composed of layers of SrVO$_3$ (SVO) alternating with layers of LaVO$_3$ (LVO). If we idealize the structures as cubic perovskites, then the layers alternate along the $[001]$ direction.  In bulk, SVO crystallizes in the ideal cubic perovskite structure,\cite{Rey1990101} while LVO crystallizes in a lower symmetry  $Pnma$ structure derived from the cubic perovskite via a four unit-cell pattern of octahedral tilts. \cite{Bordet1993253} The crystal structure of bulk solid solutions La$_{1-x}$Sr$_x$VO$_3$ interpolates between that of the two end-members with the rotation amplitude decreasing as $x$ increases. In the superlattice, the presence of a substrate and the breaking of translation symmetry can lead to different rotational distortions of the basic perovskite structure and also to a difference between lattice constants parallel and perpendicular to the growth direction.

Octahedral rotations in  perovskites can be described using Glazer's notation.\cite{Glazer:a09401} In the coordinate system defined by the three V-O bond directions of the original cubic perovskite, there are 3 tilt angles $\alpha,\beta$ and $\gamma$ with corresponding rotation axes $[100],[010]$ and $[001]$. The tilt is in-phase if successive octahedra rotate in the same direction, and anti-phase if they rotate in opposite directions. Rotational distortions of the cubic perovskite $AB$O$_3$ structure may be denoted by  $a^ib^jc^k$ where $i,j,k$ can be $+,-$ or $0$ denoting in-phase, anti-phase or no tilting, respectively and $a=\alpha,b=\beta,c=\gamma$.\cite{Glazer:a09401,Woodward:br0052,Woodward:br0058} Bulk LVO ($Pnma$) is of the type $a^-b^+a^-$ with $\alpha=\gamma=8.7^\circ$ and $\beta=7.9^\circ$.\cite{Bordet1993253,PhysRevB.85.184101}

For superlattices, substrate-induced strain may change the situation in a way which depends on the growth direction. Experiments\cite{PhysRevB.85.184101,PhysRevB.83.125403} confirm that the growth direction for the experimentally relevant superlattices is $[001]$ (in the ideal cubic perovskite notation) and we focus on this case here. Recent experimental studies of superlattices\cite{PhysRevB.83.125403} and of LaVO$_3$ thin films, which apparently have the same growth direction,\cite{PhysRevB.85.184101}  suggest that the rotations are of the type $a^-a^+c^-$\cite{Woodward:br0052,Woodward:br0058}  and indicate that the dominant rotation is around the axis defined by the growth direction: $\alpha=\beta\approx3^\circ$ and $\gamma\approx11.5^\circ$. This distortion pattern is different from that occurring in bulk.  To explore its effects we set $\alpha=\beta=3^\circ$ and consider the consequences of varying $\gamma$. 

In bulk La$_{1-x}$Sr$_x$VO$_3$, while the 4-sublattice $Pnma$ structure implies a difference in lattice constants, all V-O bond lengths are the same.\cite{Bordet1993253} The difference in lattice constants arises from a difference in tilting pattern. Superlattices are typically grown on a substrate, and in epitaxial growth conditions the lattice constants perpendicular to the growth direction  (which we denote here by $a$) are fixed by the substrate, while the lattice parameter along the growth direction ($c$) is free to relax. The result is a $c/a$ ratio typically $\neq 1$ contributed by both tilting and anisotropy in V-O bond lengths and possibly varying from layer to layer of the superlattice. For the experimentally studied superlattices, $c/a \sim 1.02$. \cite{PhysRevB.80.241102,PhysRevB.83.125403} The V-O bond lengths have not been determined but, as discussed in more detail in the Appendix, our studies indicate that all V-O bonds have essentially the same length. Further we show that a few percent differences have no significant effect on our study of ferromagnetism. In the rest of the paper we therefore ignore these distortions, setting all V-O bond lengths to be equal.

\subsection{Electronic structure}
\begin{figure}[t]
 \includegraphics[width=\columnwidth]{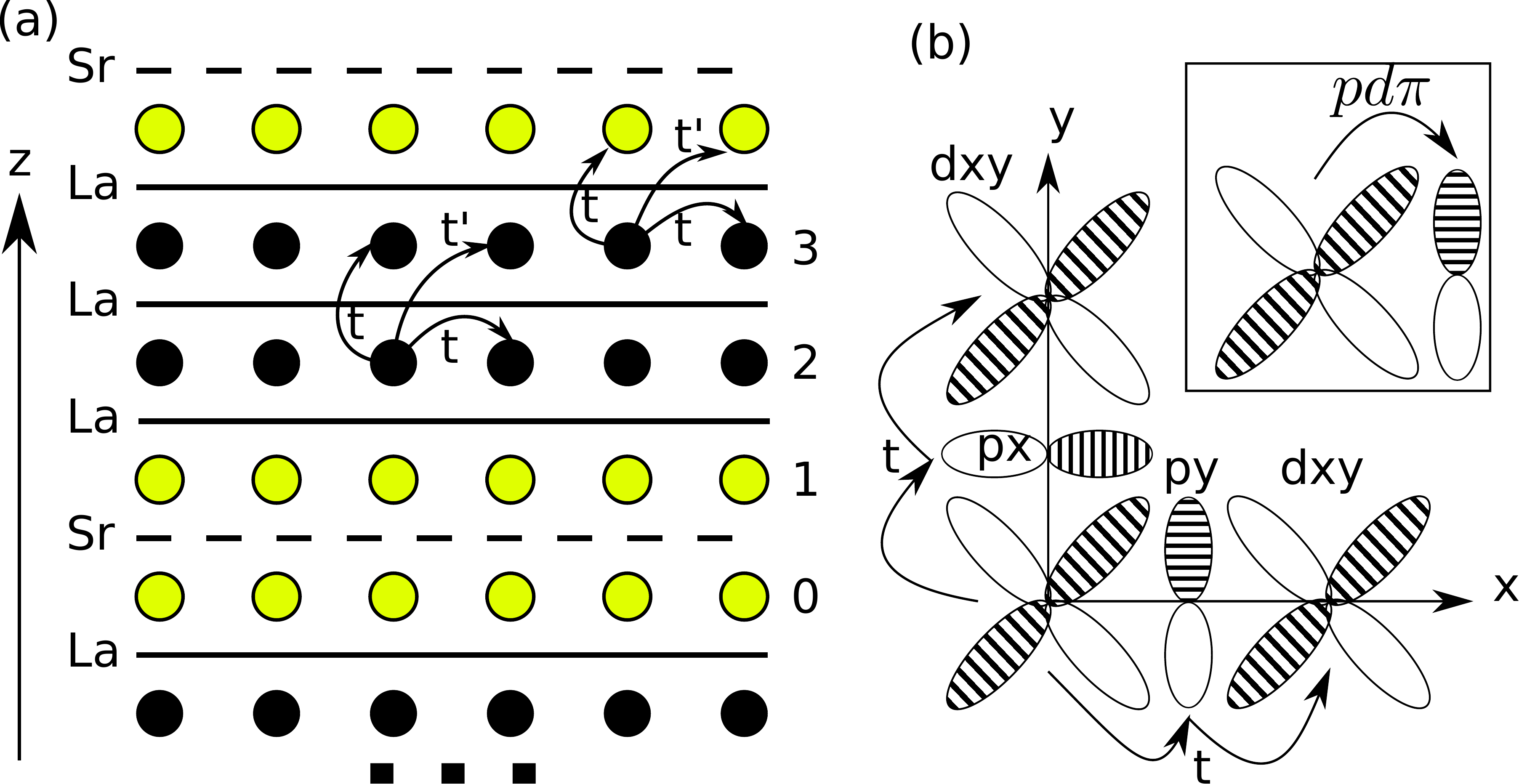}
\caption{\label{fig:cartoon}(Color online) (a) Schematic of superlattice lattice structure (LaVO$_3$)$_m$(SrVO$_3$)$_1$ with $m=3$. Vanadium sites indicated as circles with charge density indicated by shading: heavy shading (black online) indicating higher charge density and light shading (yellow online) indicating lower charge density. LaO and SrO planes are shown as solid and dashed lines respectively. Nearest neighbor ($t$) and next-nearest neighbor ($t'$)  hoppings between vanadium sites indicated by arrows. The numbers on the right are VO$_2$ layer indices. (b) Inset: $pd\pi$ hopping between $t_{2g}$ orbital and $p$-orbital. Main panel: two-dimensional nearest neighbor hopping $t$ made of two $pd\pi$ hoppings from $xy$ orbital of one vanadium site to oxygen $p_x$ or $p_y$ orbital, then to $xy$ orbital of another vanadium site.}
\end{figure}

We study superlattices designed to be similar to the system studied in Ref.~\onlinecite{PhysRevB.80.241102}. In these superlattices, units of $m$ layers of LaVO$_3$ are separated by one layer of SrVO$_3$. To define the superlattice, we begin from LVO in the appropriate bulk structure, then break translation invariance along the $[001]$ ($z$-direction) by replacing every (m+1)$^\mathrm{th}$ LaO plane with an SrO plane. Fig.~\ref{fig:cartoon}a shows such a superlattice with $m=3$. 

We assume that the superlattice is grown epitaxially so that in-plane bond lengths and other aspects of the local structure including rotations are the same for all layers.  We therefore take the electron transfer integrals which define the band structure to the be same for all layers. In this case the electronic structure of a superlattice is  defined by adding the electrostatic potentials of the Sr and La ions to the basic translationally invariant hopping Hamiltonian describing the bulk materials.

In our calculations we follow the common practice in studies of early transition metal oxides by assuming that the energy splitting between transition metal $d$-bands and oxygen $p$-bands is large enough to justify the use of a  ``frontier orbital'' model focusing on the $p$-$d$ antibonding bands which are mainly composed of vanadium $t_{2g}$-symmetry $d$-states.

The Hamiltonian for the superlattice is thus
\begin{equation}
 H = H_{kin} + H_{onsite} + H_{coulomb},
\label{eqn:fullH}
\end{equation}
where $H_{coulomb}$ describes the electron-ion interaction and electron-electron interaction between different sites and $H_{onsite}$ describes the $d$-$d$ interactions, which we take to be on-site.
 $H_{kin}$ is a tight binding model, derived by using maximally-localized Wannier function (MLWF) techniques\cite{PhysRevB.56.12847}   to fit the $t_{2g}$-derived antibonding bands. The detailed procedure is described in our previous work.\cite{PhysRevB.87.155127}  

The kinetic Hamiltonian has the quadratic form 
\begin{equation}\label{eqn:Hkin}
H_{kin} = \sum_{\mathbf{k},\alpha,\beta,\sigma} H_{band}^{\alpha\beta}(\mathbf{k})c^\dagger_{\mathbf{k}\alpha\sigma} c_{\mathbf{k}\beta\sigma},
\end{equation}
where $c^\dagger_{\mathbf{k}\alpha\sigma}$ and $c_{\mathbf{k}\beta\sigma}$ are electron creation and annihilation operators in reciprocal space with wavevector $\mathbf{k}$. $\alpha$ and $\beta$ are orbital and layer indices, and $\sigma$ is the spin index.

We assume that the interaction takes the standard Slater-Kanamori form \cite{PhysRev.49.537,Kanamori195987,PTP.30.275}  which following Ref.~\onlinecite{PhysRevB.87.155127} we write as

\begin{equation}\label{eqn:Honsite}
\begin{split}
H_{onsite} = &U\sum_{i\alpha}n_{i\alpha\uparrow}n_{i\alpha\downarrow}  + (U-2J)\sum_{i\alpha\neq \beta}n_{i\alpha\uparrow}n_{j\beta\downarrow} + \\
        & + (U-3J)\sum_{i,\alpha > \beta,\sigma}n_{i\alpha\sigma}n_{i\beta\sigma} + \\
        & + J\sum_{i\alpha\neq \beta}\psi^\dagger_{i\alpha\uparrow}\psi_{i\beta\uparrow}\psi^\dagger_{i\beta\downarrow}\psi_{i\alpha\downarrow} + \\
        & + J\sum_{i\alpha\neq \beta}\psi^\dagger_{i\alpha\uparrow}\psi_{i\beta\uparrow}\psi^\dagger_{i\alpha\downarrow}\psi_{i\beta\downarrow},
\end{split}
\end{equation}
where the values of the on-site interaction $U$ and the Hund's coupling $J$ are $U=6\mathrm{eV}\sim22t$ and $J=1$eV so that LVO is an insulator in bulk while SVO is a metal.

In the approximation employed here, the superlattice is defined by  the Coulomb interaction between the La/Sr ions and electrons. This, and the off-site part of the  electron-electron interaction is contained \cite{PhysRevB.70.075101} in 
\begin{equation}
 H_{coulomb} = H_{el-ion} + H_{el-el}.
\label{eqn:Hcoulomb}
\end{equation}
To construct $H_{el-ion}$, we assume that the whole ion charge of SVO or LVO unit cell comes into the Sr or La site. Consider SrVO$_3$, the valence of V is $+4$ ($d^1)$. If this one $d$-electron is removed, the SVO unit cell will have charge $+1$, hence, in our model, Sr site has charge $+1$. Similarly, LaVO$_3$ has V$^{+3}$ ($d^2$),  thus La site has charge $+2$. As a result, $H_{el-ion}$ has the form
\begin{equation}
\begin{split}
 H_{el-ion} &= \sum_{i, R_{Sr}}\dfrac{-e^2 \hat{n}_i} {4\pi\epsilon\epsilon_0|R_i - R_{Sr}|} + \\
     &+\sum_{i, R_{La}} \dfrac{-2e^2 \hat{n}_i} {4\pi\epsilon\epsilon_0|R_i - R_{La}|}.
\end{split}
\label{eqn:Hel-ion}
\end{equation}
where $n_i$ is electron-occupation operator at V-site $i$, $\epsilon$ is the relative dielectric constant. The part $H_{el-el}$ is the inter-site Coulomb interaction of vanadium $d$-electrons
\begin{equation}
 H_{el-el} = \dfrac{1}{2}\sum_{\substack{i,j \\ i \ne j}}\dfrac{e^2 \hat{n}_i n_j} {4\pi\epsilon\epsilon_0|R_i - R_j|}.
\label{eqn:Hel-el}
\end{equation}

$H_{el-el}$ is treated in the Hartree approximation. Note that in Eq.~\eqref{eqn:Hel-el}, $\hat{n}_i$ is the operator giving the total $d$-electron occupation of site $i$, while $n_j=\langle\hat{n}_j\rangle$ is the expectation value of $d$-electron occupancy at site $j$, which is determined self consistently. From $H_{coulomb}$, the Coulomb potential $V_i$ for site $i$ is calculated using Ewald summation.\cite{Ewald_sum}

The dielectric constant $\epsilon$ is an important parameter in Eqs.~(\ref{eqn:Hel-ion},~\ref{eqn:Hel-el}). It accounts for screening on the scale of a lattice constant so bulk measurements are not directly relevant and  an appropriate value has not been determined. Values ranging from $4$ to $15$ have been reported in the literature for similar systems.\cite{PhysRevLett.97.056802,PhysRevLett.101.227004}  Because the appropriate value of $\epsilon$ has not been determined, we have studied several cases and  present results mainly for $\epsilon = 8, 15$.

\subsection{Methods}
We treat the on-site interaction terms using single-site dynamical mean field theory (DMFT)\cite{Georges96}  with the hybridization expansion continuous time quantum Monte Carlo (CTQMC) solver.\cite{PhysRevLett.97.076405} The superlattice effect is taken into account by the Coulomb potential $\hat{V}$. We use the  superlattice dynamical mean field theory introduced by Potthoff and Nolting \cite{PhysRevB.59.2549,PhysRevB.67.165408} in the form given in Ref.~\onlinecite{PhysRevB.70.075101}. Here  each V site $i$ has a self energy (site local but dependent on site) determined from the solution of a quantum impurity model which has parameters fixed by the DMFT self-consistency equation linking the site local term of the lattice Green function $\{\}_{ii}$ to the quantum impurity model Green function\cite{code_download}
\begin{equation}
 \hat{G}^i_{imp}(\omega) = \left\{\left[(\omega+\mu)\mathbb{1} - \hat{H}_{band} - \hat{V} - \hat{\Sigma}(\omega) \right]^{-1}\right\}_{ii},
\end{equation}
where
\begin{equation}
\begin{split}
 V_i = & \sum_{\substack{j, j \ne i}}\dfrac{e^2 n_j} {4\pi\epsilon\epsilon_0|R_i - R_j|} - \sum_{R_{Sr}}\dfrac{e^2} {4\pi\epsilon\epsilon_0|R_i - R_{Sr}|} - \\
       & -\sum_{R_{La}} \dfrac{2e^2} {4\pi\epsilon\epsilon_0|R_i - R_{La}|}
\end{split}
\end{equation}
is a  site dependent quantity, diagonal in spin and orbital indices but linking different sites, derived from Eqs.~(\ref{eqn:Hel-ion},~\ref{eqn:Hel-el}). The layers are coupled by a self-consistency condition which as discussed in Refs.~\onlinecite{PhysRevB.59.2549,PhysRevB.67.165408,PhysRevB.70.075101} fixes both the hybridization function of the quantum impurity model and the layer-to-layer variation in the  charge density.

As described in Ref.~\onlinecite{PhysRevB.87.155127}, it is advantageous to perform a site-local rotation to align the orbital basis to the local V-O bond directions of  each octahedron before solving the impurity model. This reduces  the sign problem in the CTQMC impurity solver and restores in-plane translation invariance in the sense of making the self-consistency equations the same for all sites in a given plane. 

In a superlattice composed of $N$ layers, it is in principle necessary to solve $N$ dynamical mean field problems, coupled by the self-consistency condition. However, we find (see section \ref{sec:cubicsuperlattice}) that the susceptibility for a given layer of the  superlattice may be determined from a bulk computation at the same local density and crystal structure. Because the layer dependent density has no significant dependence on the temperature or the many-body physics, it may be determined once from a band structure calculation and then bulk results with the appropriate density for a wide range of temperature may be used to infer the Curie temperature, substantially reducing the computational burden.

The Curie temperature for ferromagnetism is determined by extrapolating the inverse susceptibility $\chi^{-1}(T)$ to $0$ based on Curie-Weiss law $\chi^{-1} \sim T-T_c$. The test for the reliability of this method for $T_c$ has been done in Ref.~\onlinecite{PhysRevB.87.155127}. A similar approach can be found in literature.\cite{PhysRevB.78.205118}

\section{Relation between superlattice and bulk system calculations\label{sec:cubicsuperlattice}}
In this section, we demonstrate that the magnetic phase diagrams of superlattice systems may be inferred, to reasonable accuracy, from the study of appropriately chosen bulk systems. This enables a considerable reduction in the computation resources required. 

\begin{figure}[h]
 \includegraphics[width=\columnwidth]{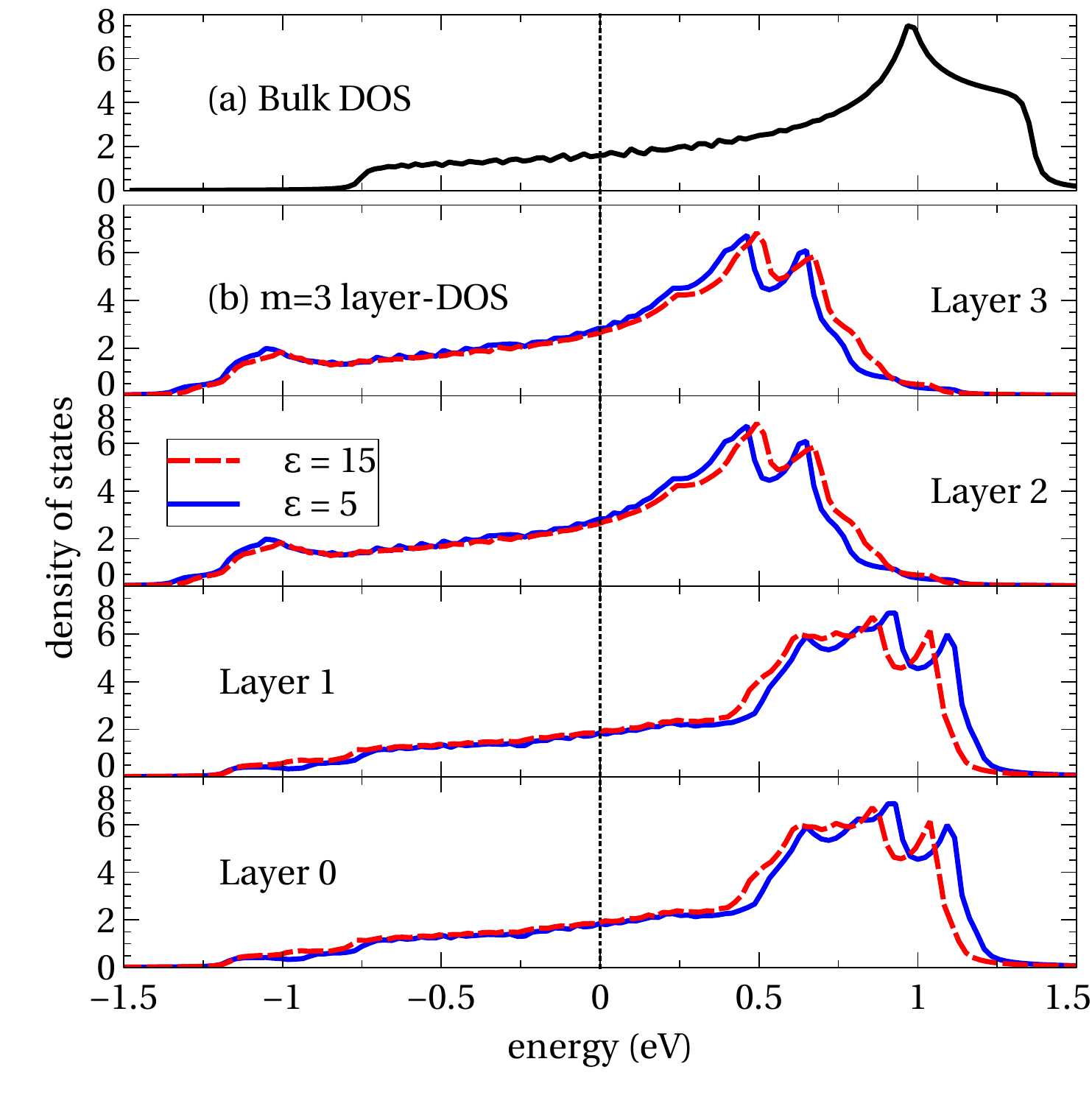}
\caption{\label{fig:DOS}(Color online) Panel (a): Non-interacting density of states for bulk system at carrier density $n=1$. Panels (b): Non-interacting density of states for different layers of (LVO)$_3$(SVO)$_1$ superlattice for two different values of dielectric constant $\epsilon = 5$ (solid) and $\epsilon = 15$ (dashed) with hopping parameters $t = 0.264$eV and $t' = 0.084$eV. SrO plane is between layers $0$ and $1$ (the index is defined in Fig.~\ref{fig:cartoon}). The Fermi energy is at $0$.}
\label{nonintdos}
\end{figure} 

We begin with a  study of  ``untilted'' or ``cubic'' superlattices: those in which  all V-O-V bond angles are $180^\circ$. We  focus specifically on  $[001]$ superlattices in which the unit cell contains  $m$ layers LVO and one layer SVO, where $m = 3, 4, 5$. For orientation, we present the density of states (DOS) of the non-interacting system in Fig.~\ref{fig:DOS}.  In obtaining these densities of states we used the simple tight binding parametrization. The  DOS for the bulk system is shown  in panel (a). One sees the typical three-fold degenerate DOS for $t_{2g}$ band, the Van Hove singularity is visible as a peak near the upper band edge. It is at high energy because the next-nearest neighbor hopping $t' > 0$. The remaining panels show the  layer-resolved densities of states for the $m=3$ superlattice. The upper two panels show layers sandwiched by La on both sides; the lower two panels show the layers adjacent to the SrO plane. The superlattice-induced changes in the density of states are seen to be relatively minor:  the main effects are  a weak splitting of the van Hove peaks reflecting the breaking of translational invariance in the $z$-direction, and a relative shift in the positions of the van Hove peaks arising from band bending associated with the different charges of the Sr and La ions.

\begin{figure}[ht]
 \includegraphics[width=\columnwidth]{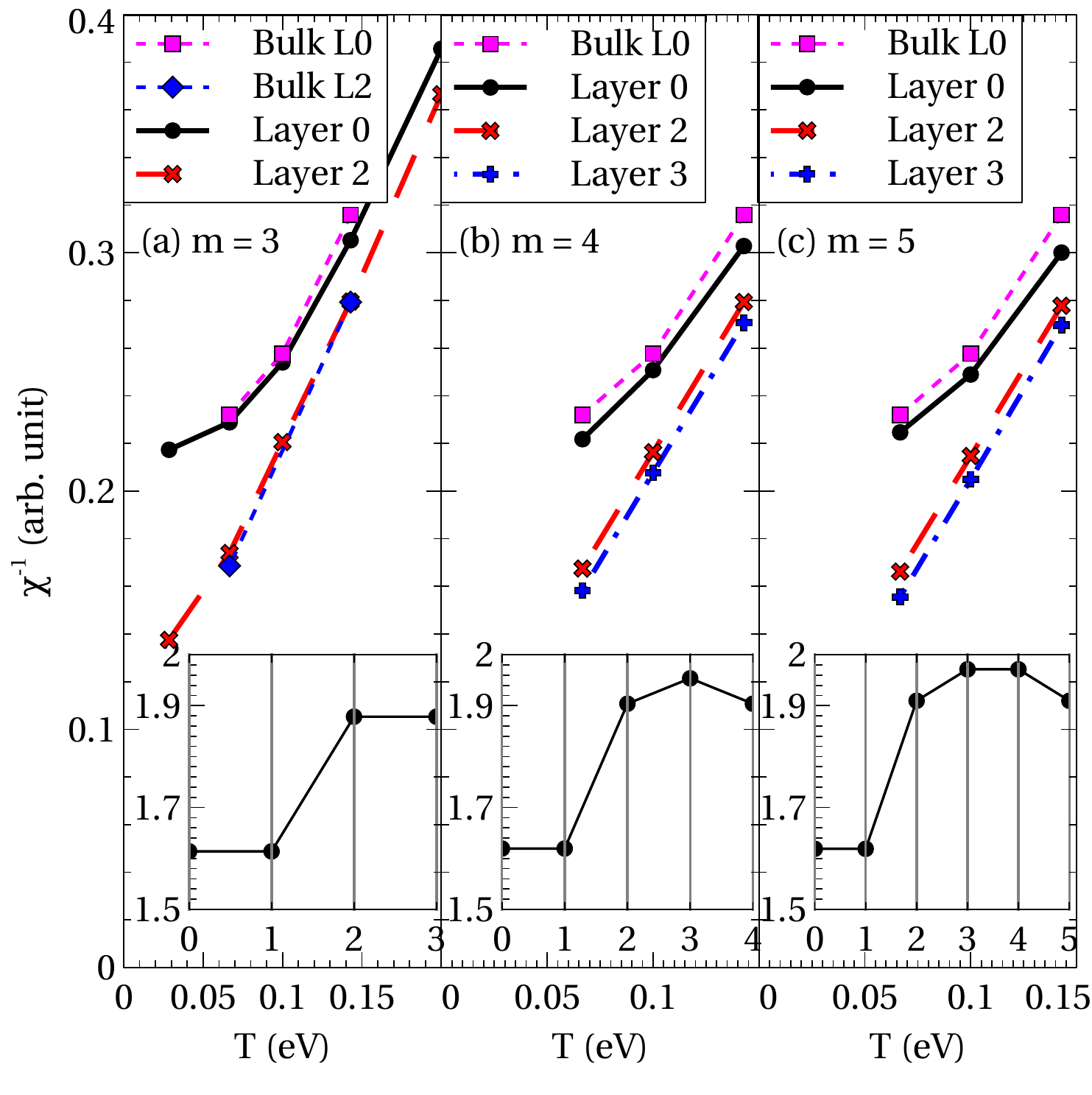}
\caption{\label{fig:SLchi}(Color online) Temperature-dependent layer-resolved inverse magnetic susceptibilities for symmetry-inequivalent layers of untilted (LVO)$_m$(SVO)$_1$ superlattice structures with different numbers of LVO layers $m=3,4$ and $5$. Layer $0$ is adjacent to SrO and layers $2$ and $3$ are between two LaO layers. The relative dielectric constant is $\epsilon = 15$, magnetic field $H = 0.01\mathrm{eV}/\mu_B$. The $\chi^{-1}(T)$ obtained from solution of bulk cubic systems with charge density set to the density on the given layer are also shown. ``Bulk L0'' (``BulkL2'') denotes a calculation performed for a bulk system with density the same as for $L=0$ ($L=2$) layer density. Inset: the electron layer density distribution corresponding to the susceptibility plot, $x$-axis is the layer index, $y$-axis is the layer density. On-site interactions $U=6$eV, $J=1$eV.}
\end{figure}

Fig.~\ref{fig:SLchi} shows the layer-resolved charge density and inverse susceptibility $\dfrac{H}{m(H)}$ plotted against temperature for three different superlattice structures corresponding to $m=3,4,5$. As expected from electrostatic considerations, the charge is lower for the VO$_2$ planes nearer the SrO layer and the charge variation between layers is controlled by the dielectric constant. 

The magnetization $m$ at the V sites on each layer was computed at field $H = 0.01\mathrm{eV}/\mu_B$ and the inverse susceptibility was obtained as $H/m$. Linearity was verified by repeating the computation using  $H = 0.02\mathrm{eV}/\mu_B$ (not shown). For the $m=3, \epsilon=15$ case (Fig.~\ref{fig:SLchi}a), we extended the computation to the lower temperature $T=0.03$eV; for the other two cases $T=0.06$eV was the lowest temperature studied.  The inverse susceptibilities are approximately linear in temperature at higher temperatures and in all cases, extrapolation to $\chi^{-1}=0$ reveals $T_c<0$, implying absence of ferromagnetism. 

Especially for the layer nearest the SrO plane the $\chi^{-1}$ curves  exhibit weak upward  curvature at the lowest temperatures studied. As shown in Ref.~\onlinecite{PhysRevB.87.155127}, the curvature is a signature that the system is entering a Fermi-liquid coherence regime. The Fermi liquid coherence temperature is highest for the layers nearest the SrO because the charge in these planes is farther from the $n=2$ Mott insulating state.  To verify this we followed Ref.~\onlinecite{PhysRevB.87.155127} and computed the Wilson ratio $R_W$  for each layer of the superlattice for the case $m=3,\epsilon=15$, finding (not shown) that for each layer the $R_W$ extrapolates to $2$ at low temperature. The approach to the low temperature value is faster for layers with low density (near SrO planes) than for layers with high density (far from SrO planes). $R_W=2$ is the value for a Kondo lattice, while ferromagnetism is characterized by an $R_W>2$.\cite{PhysRevB.87.155127} We therefore believe that for ``untilted'' superlattices, the differences in $\chi^{-1}$ among layers arise from differences in quasiparticle coherence scale, there is no evidence for ferromagnetism in this system, consistent with the solution of the corresponding bulk problem.

To gain insight into the physics underlying the layer dependence of $\chi^{-1}$ we have computed $\chi^{-1}(T)$ for the cubic bulk system ($H_{coulomb}=0$, $H_{kin}$ is constructed from the two-dimensional dispersion $\epsilon(\mathbf{k})=-2t(\cos k_x+\cos k_y)-4t'\cos k_x\cos k_y$) for carrier densities equal to those on the different VO$_2$ layers. In Fig.~\ref{fig:SLchi}, we present bulk calculations for $n=1.62$ and $n=1.88$ corresponding to the  densities calculated for layer $0$ and $2$ of the superlattice for all cases $m=3,4,5$. For $n=1.88$, bulk $\chi^{-1}$ at $T=0.06, 0.10$ and $0.14$eV are very close to those of $L=2$ layer of $m=3$ superlattice, which has the same density. For $m=4,5$ superlattices, bulk $n=1.88$, $\chi^{-1}(T)$ (not shown) almost coincides with those of $L=2$ layer. For bulk $n=1.62$, the difference between bulk and superlattice $L=0$ layer is small. These calculations demonstrate a general rule: within the single-site DMFT approximation, the layer-resolved properties of a superlattice correspond closely to those of the corresponding bulk system at a density equal to that of the superlattice.

\begin{figure}[h]
 \includegraphics[width=\columnwidth]{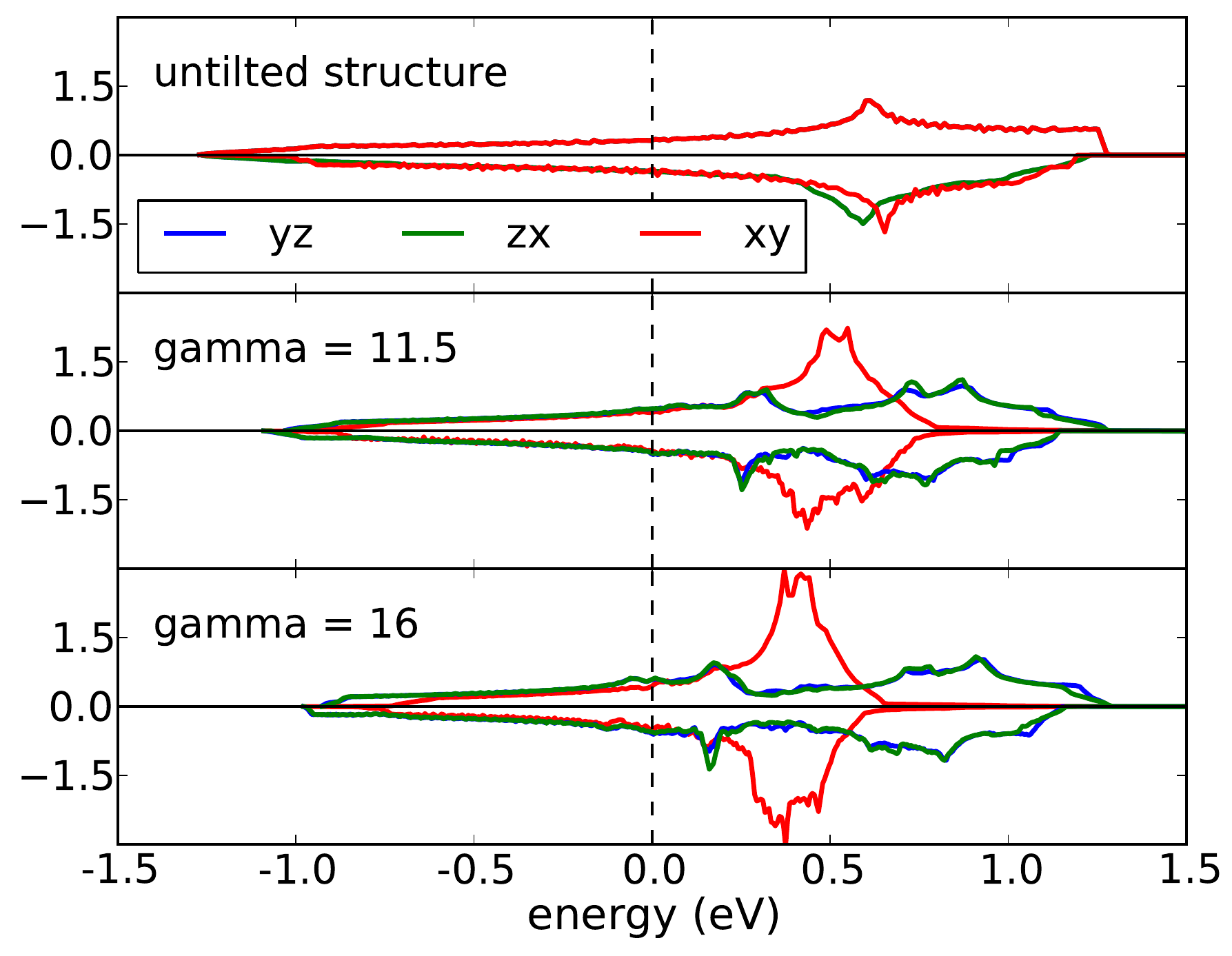}
\caption{\label{fig:DOS_bulk_sl}(Color online) Comparison between bulk LVO partial DOS (positive curves) and (LaVO$_3$)$_3$(SrVO$_3$)$_1$ superlattice layer DOS of layers near SrO (negative curves) derived from band structure calculations (DFT+MLWF). Both systems have the same lattice structure for each case: untilted structure for the top panel and $P2_1/m$ structure (Glazer's notation $a^-a^+c^-$) with $\alpha=\beta=a=3^\circ$ and $\gamma=c=11.5^\circ$ and $16^\circ$ for other panels. The DOS of bulk system is shifted towards higher energy so that bulk carrier density is the same as layer density of superlattices for the layers near SrO ($n\approx1.55$). The vertical dashed line marks the Fermi level.}
\end{figure} 

The superlattices of experimental relevance have crystal structures which are distortions of the ``untilted'' one, involving in particular a $P2_1/m$ structure characterized by a rotational distortion of the $a^-a^+c^-$ type\cite{Woodward:br0052,Woodward:br0058} involving a  large rotation about an axis approximately parallel to the  growth direction and much smaller rotations about the two perpendicular axes. Fig.~\ref{fig:DOS_bulk_sl} compares the non-interacting DOS of bulk and  (LVO)$_3$(SVO)$_1$ superlattice systems (both with the same $P2_1/m$ structure)  calculated using DFT and a MLWF parametrization of the frontier bands. The DOS of bulk system is shifted so that it has the same carrier density as layers of the superlattices near SrO plane. For three different structures (untilted structure and $P2_1/m$ structure with $\gamma=11.5^\circ$ and $16^\circ$), the basic features of the partial DOS are similar between bulk and superlattice. The translation symmetry breaking in $z$-direction leads to small extra peaks in the superlattice DOS. These differences are smoothed out by the large imaginary part of the DMFT self energy. Because the DMFT equations depend only on the density of states it is reasonable to expect that, as in the untilted case, they  will therefore give the same results in the superlattice as in the bulk material with corresponding density of states. 

\begin{figure}[h]
 \includegraphics[width=\columnwidth]{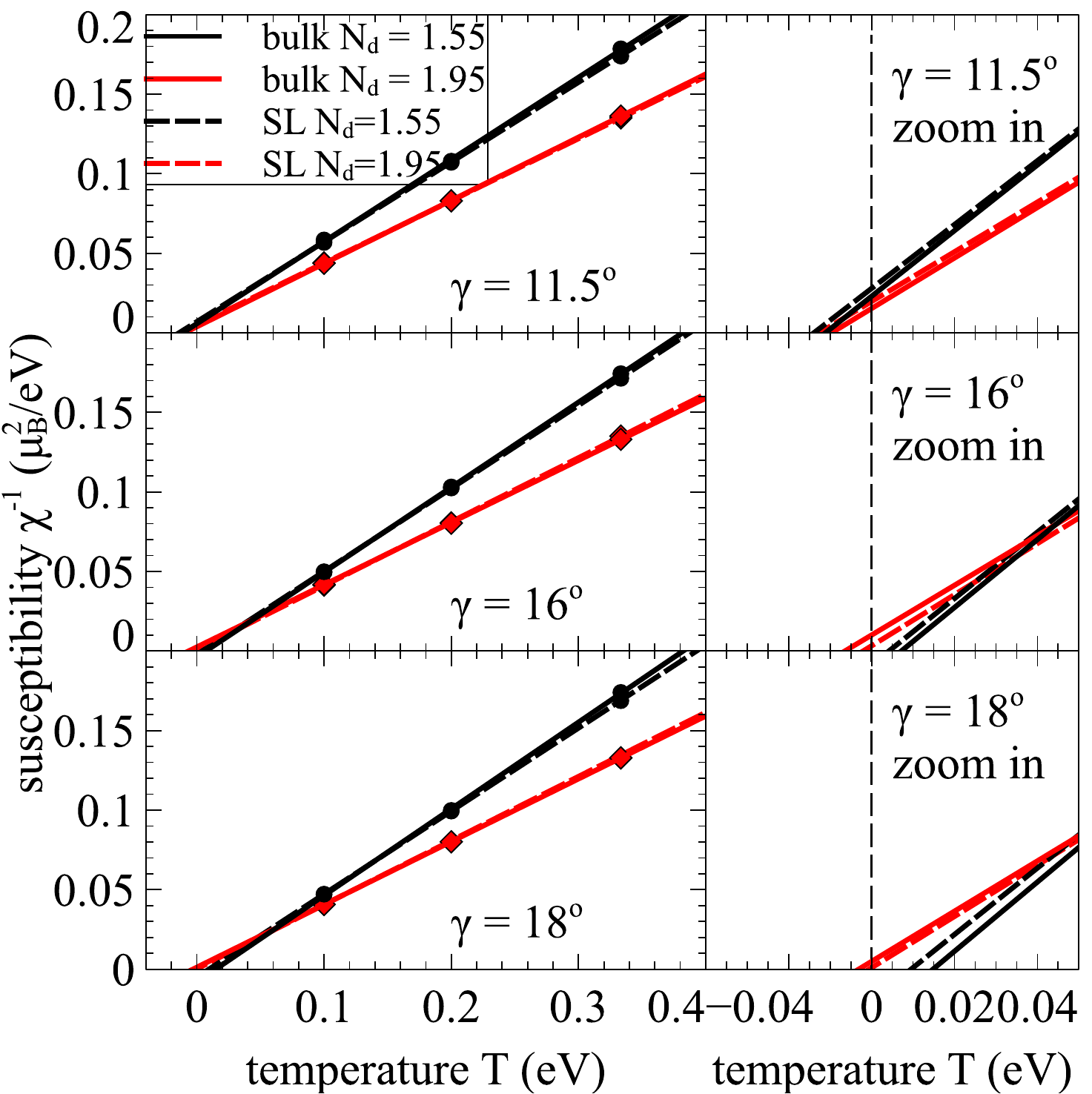}
\caption{\label{fig:bulkvslayertilted}(Color online) Comparison in temperature dependent inverse susceptibility between bulk LVO (solid lines) and (LaVO$_3$)$_3$(SrVO$_3$)$_1$ superlattice (dashed lines). Both have the same lattice structure $P2_1/m$ with tilt angle $\gamma=11.5,16$ and $18^\circ$. Bulk system has the same densities as those of layers of superlattice near and far from SrO planes ($n=1.55,1.95$). Left column: the plots in wide temperature range. Right column: the expanded views near zero temperature.}
\end{figure}

To verify that this is the case we have also compared bulk and superlattice susceptibilities for tilted structures. The four VO$_6$ octahedra in a unit cell are related by rotation, so an appropriate choice of local basis means that only one calculation needs to be carried out for a given layer. Fig.~\ref{fig:bulkvslayertilted}  compares the inverse susceptibilities for an $m=3$ superlattice to calculations performed on a bulk system with the same $P2_1/m$ structure. In these calculations, we choose $\gamma=11.5^\circ, 16^\circ, 18^\circ$ and dielectric constant $\epsilon = 8$. We see that in this case, as in the ``untilted'' case, the superlattice inverse susceptibilities $\chi^{-1}(T)$ are almost the same as those for bulk system calculated at the same density, with differences only resolvable in the expanded view for the largest tilt angles.

\section{Superlattices with $\mathrm{GdFeO}_3$-type rotation\label{sec:tiltedsuperlattice}}

In this section we present and explain our results for the magnetic phase diagram of  (LVO)$_m$(SVO)$_1$ superlattices with  the $P2_1/m$ structure (Glazer's notation $a^-a^+c^-$) reported for the experimental systems.\cite{PhysRevB.85.184101,PhysRevB.83.125403} In these structures  in-plane rotation along the growth direction $\hat{z}=[001]$ is large $\gamma=c=11.5^\circ$ (presumably because of the strain imposed by the substrate), while the out-of-plane rotation is small ($\alpha=\beta=a=3^\circ$) perhaps because the system is free to relax along the growth direction.  We concentrate on the effect of the large rotation by fixing the in-plane angles to  $3^\circ$ while varying the out-of-plane angles over a wide range from $10^\circ\to18^\circ$.

\begin{figure}[h]
 \includegraphics[width=\columnwidth]{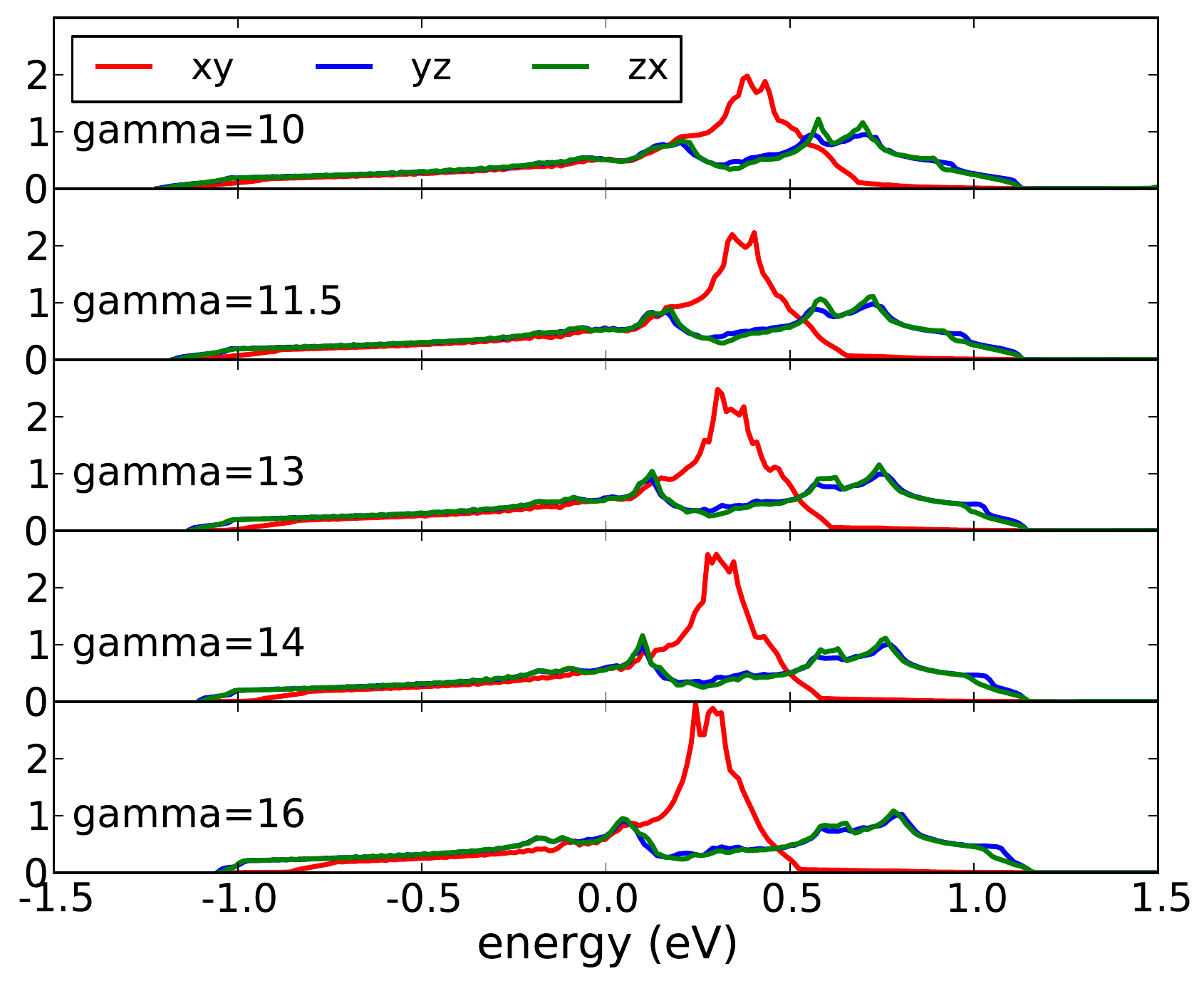}
\caption{\label{fig:pdos_P21m}(Color online) Partial DOS derived from DFT+MLWF for ``bulk'' $P2_1/m$ structure (Glazer's notation $a^-a^+c^-$) with $\alpha=\beta=a=3^\circ$ and $\gamma=c$ changing from $10^\circ$ to $16^\circ$. Only $t_{2g}$ bands are plotted because $e_g$ bands are negligible in this range of energy.}
\end{figure} 

Based on the results of Section ~\ref{sec:cubicsuperlattice} we generate  a phase diagram for the superlattice from calculations for a bulk system which is a $P2_1/m$ distortion of the ideal cubic perovskite structure of chemical composition LaVO$_3$.  The bulk system results are presented as a phase diagram in the plane of carrier concentration and $\gamma$-rotation. Specific layers of the superlattice will correspond to particular points on the phase diagram, with the layer dependent density fixed by number of LVO layers $m$ and the dielectric constant $\epsilon$ and the rotation fixed by the substrate lattice parameter.

We use DFT+MLWF methods to obtain the frontier orbital band structure for the $t_{2g}$-derived antibonding bands.   Fig.~\ref{fig:pdos_P21m} presents representative results for the orbitally resolved local density of states.  In this figure the orbitals are defined with respect to the local basis defined by the 3 V-O bonds of a given VO$_6$ octahedron. We define  $\hat{z}=[001]$ as the axis (approximately parallel to the growth direction) about which the large rotation occurs.  Fig.~\ref{fig:pdos_P21m} shows that $yz$  and $zx$ orbitals are almost degenerate, while $xy$ orbital is strikingly different.   The DOS of $xy$ orbital maintains the shape of a two-dimensional energy dispersion with a van Hove peak well above the chemical potential, similar to the bulk cubic structure (see e.g. Fig.~\ref{fig:DOS}a). There are noticeable differences only at very high rotation angles. On the other hand, $yz$ and $zx$ orbitals are spread out with two small peaks, because hoppings along $x$ or $y$ directions (more distorted) are different from those along $z$-direction (less distorted). When the distortion gets larger, the $d$-bandwidth becomes smaller, the $xy$ peak gets larger and slightly closer to the Fermi level, and $yz$ and $zx$ peaks near the Fermi level also develop.

\begin{figure}[h]
 \includegraphics[width=\columnwidth]{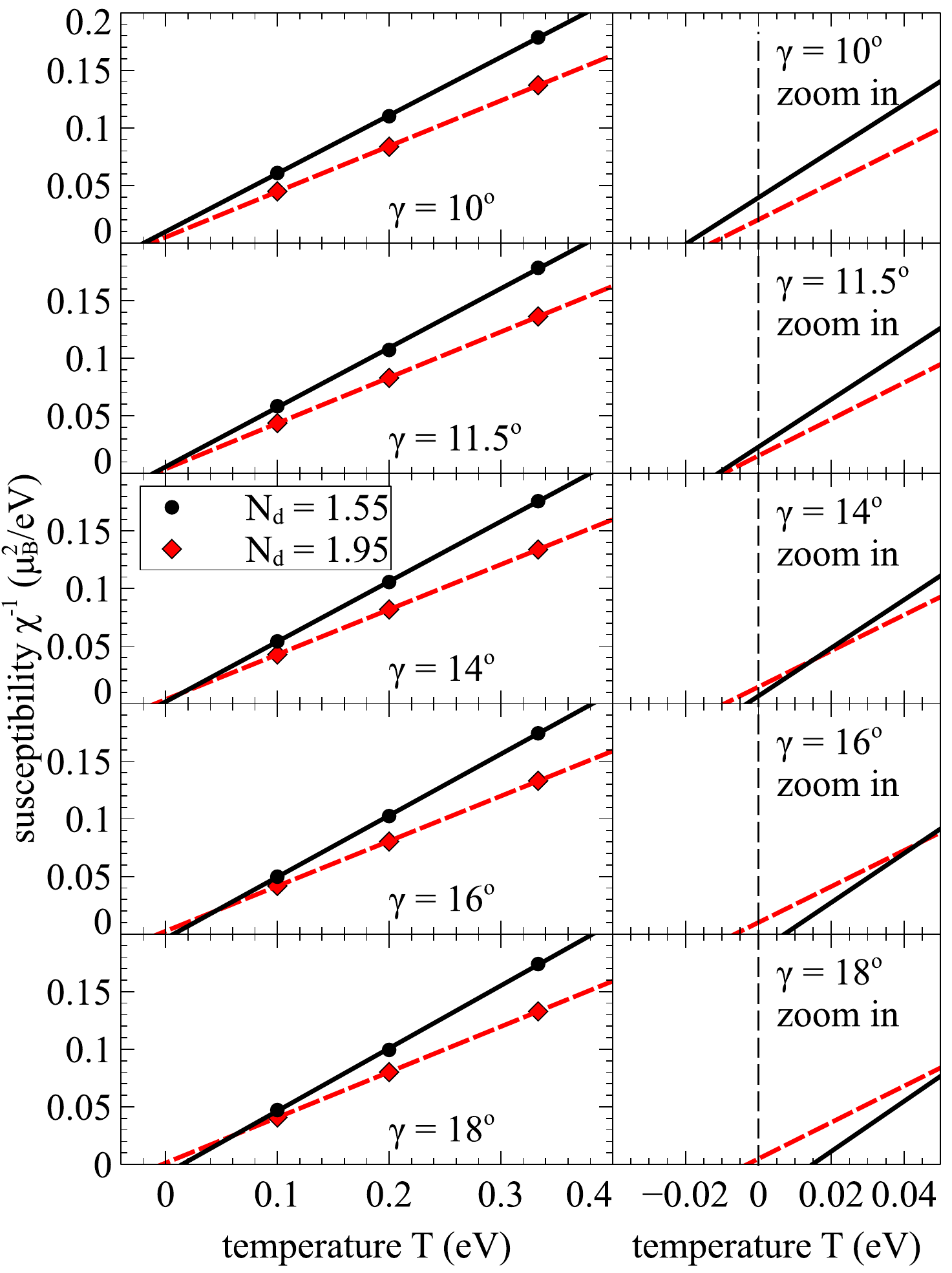}
\caption{\label{fig:Tc_evolve}(Color online) Inverse susceptibility $\chi^{-1}$ vs. temperature $T$ for bulk $P2_1/m$ structure of LaVO$_3$ at densities $n=1.55$ (black circle solid lines) and $n=1.95$ (red diamond dashed lines) for rotation angle $\gamma$ increasing from $10\to18^\circ$. On-site interaction $U=6$eV and $J=1$eV. Left column: the circles and diamonds are data points, the solid and dashed lines are fitted from these data points. Right column: expanded view at small $\chi^{-1}$ region. The vertical dashed line marks zero temperature.}
\end{figure}

Based on $\hat{H}_{band}(\mathbf{k})$ generated by DFT+MLWF, we carry out DMFT calculations for in-plane rotation angle $\gamma$ to get $\chi^{-1}$ curves whose extrapolations define the Curie temperatures $T_c$. Fig.~\ref{fig:Tc_evolve} shows how $T_c$ evolves when the rotation angle $\gamma$ increases from $10$ to $18^\circ$. In this figure, we consider two different carrier densities $n=1.55$ and $1.95$, corresponding to the band structure prediction for the layer densities  of layers near and far from SrO planes in the superlattice. $T_c$ for $n=1.95$ is a slow function of rotation and is always negative for the range of $\gamma$ under consideration, while $T_c$ for $n=1.55$ increases faster, so that the system becomes ferromagnetic when $\gamma$ is between $14$ and $16^\circ$.  Ferromagnetism is therefore expected only in superlattices with very large rotations, and then only in the layers with large hole doping (i.e. the layers closest to the SrO planes).

\begin{figure}[h]
 \includegraphics[width=\columnwidth]{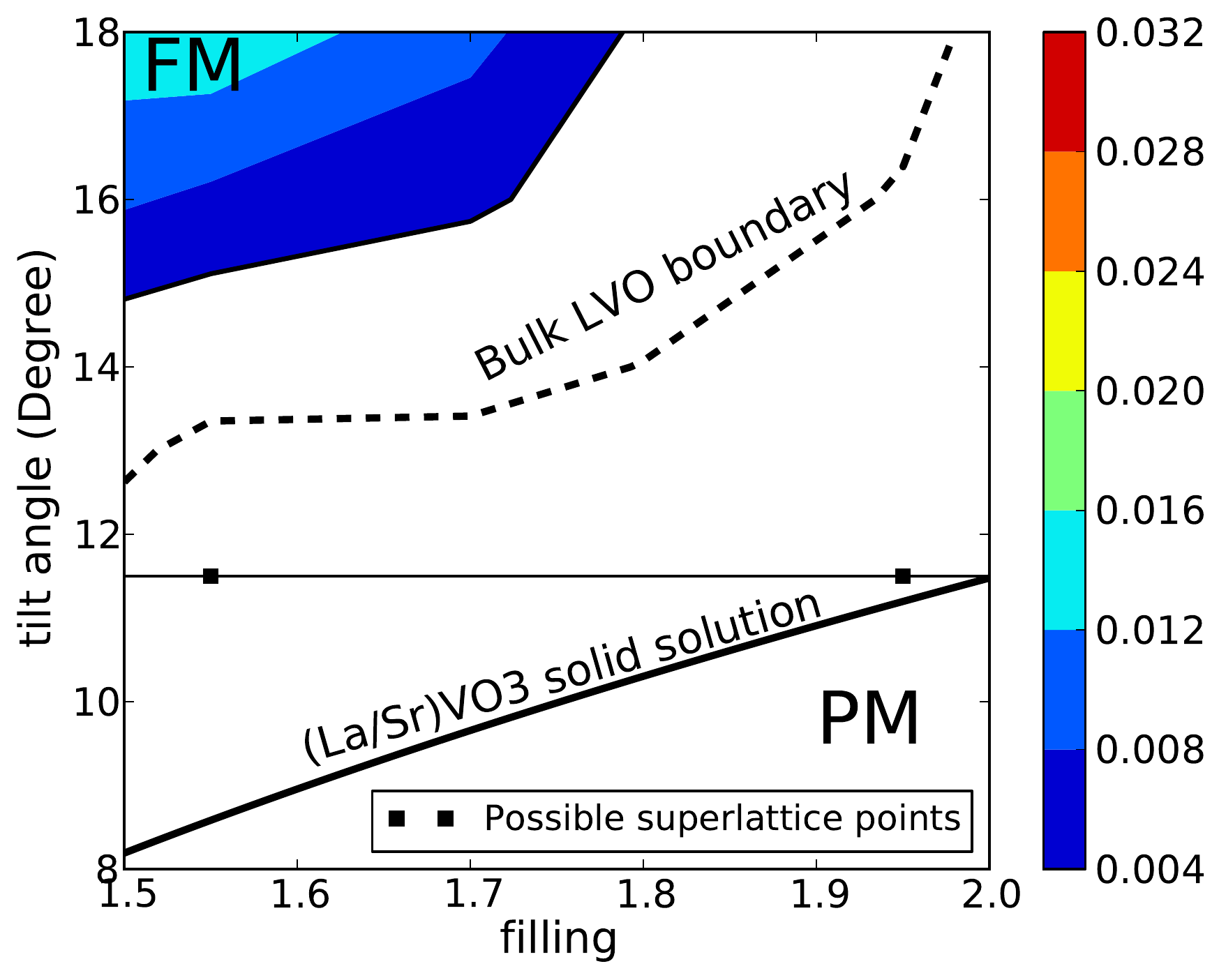}
\caption{\label{fig:phase_diagram}(Color online) The magnetic phase diagram with $x$-axis carrier density $n$ and $y$-axis tilt and rotation angle along $\hat{z}=[001]$ direction $\gamma$ for bulk system LVO with the same type of distortion as for (LVO)$_m$(SVO)$_1$ superlattices ($P2_1/m$ structure), in-plane tilt angles $\alpha,\beta\approx 3^\circ$. On-site interactions $U=6$eV, $J=1$eV.  The white regime indicates absence of ferromagnetism  ($T_c < 0.004eV$), the colored  regime indicates ferromagnetism with $T_c$ indicated by the color bar. Also indicated are results for bulk La$_{1-x}$Sr$_x$VO$_3$ in the $Pnma$ structure, from Ref.~\onlinecite{PhysRevB.87.155127}. Note that in the calculations for the $Pnma$ structure all three tilt angles are almost the same.}
\end{figure}

From a range of calculations such as those shown in Fig.~\ref{fig:Tc_evolve} we have constructed the superlattice magnetic phase diagram shown in Fig.~\ref{fig:phase_diagram}. Similar to Ref.~\onlinecite{PhysRevB.87.155127}, there are uncertainties in our extrapolation for Curie temperature, we consider $0.004$eV as the error bar for positions on the phase diagram. Thus, $T_c < 0.004$eV is considered as $T_c = 0$ within the error bar.  We see that ferromagnetism is favored only for very large rotations, much larger than the $11^\circ$ determined experimentally, and only for carrier concentrations far removed from $n=2$. We may locate the experimentally studied superlattices on this phase diagram. For an $m=3$ superlattice, band structure calculations indicate layer densities  $1.55$ for layers near SrO plane and $1.95$ for the other layers. The experimentally determined rotation angle is $\sim 11.5^\circ$.  These two points are indicated by squares in Fig.~\ref{fig:phase_diagram}.

It is interesting to compare our results to those previously obtained \cite{PhysRevB.87.155127} for the  bulk solid solution La$_{1-x}$Sr$_x$VO$_3$ ($Pnma$ structure). The dashed line in Fig.~\ref{fig:phase_diagram} shows the theoretically estimated phase diagram for the bulk solid solution. We see that the bulk structure is more favorable for ferromagnetism than the superlattice structure.  An important difference between the $Pnma$ structure and the $P2_1/m$ of the superlattice is that in the former case all three tilt angles are of comparable magnitude whereas in the $P2_1/m$ structure only one rotation is large. We believe that this difference is responsible for the difference in phase boundary.

\section{Conclusions\label{sec:conclusions}}

In this paper, we have studied the possibility of ferromagnetism in superlattice structures of vanadium oxides derived from LaVO$_3$ and SrVO$_3$. Our investigation was based on the idea that ferromagnetism depends on an interplay between carrier density and octahedral rotation, and while these are coupled in bulk (see the solid solution curve in  Fig.~\ref{fig:phase_diagram}) they may be decoupled in the superlattice. In particular, the charge density varies across the superlattice, being lowest near the SrO planes, while the rotation angle is controlled by the substrate. Thus in an appropriately designed superlattice at least some portions of the system might be moved closer to (or perhaps into) the ferromagnetic region. In several important aspects this idea is consistent with calculations. We find that the local carrier density determines the local magnetic susceptibility (see section ~\ref{sec:cubicsuperlattice}) and the density/tilt angle relationship may be significantly altered (see  solid line and square points in Fig. \ref{fig:phase_diagram}). 

However,  we find that the $P2_1/m$ octahedral rotation pattern characteristic of experimentally discovered superlattices is in fact less favorable to  ferromagnetism than the $Pnma$ pattern characteristic of bulk materials (compare the phase boundaries in Fig. \ref{fig:phase_diagram}). Thus while the general idea that an appropriately designed superlattice might provide conditions favorable for ferromagnetism thereby providing a potential explanation for the remarkable experimental report of room-temperature ferromagnetism in (LaVO$_3$)$_m$(SrVO$_3$)$_1$ superlattices with $m=3,4,5,6$ by L\"uders et. al.,\cite{PhysRevB.80.241102} (even though there is no ferromagnetism in the bulk solid solution), our detailed findings are not consistent with the experimental result.  

Our results indicate that designing  ferromagnetism into a vanadate superlattice will require both large amplitude rotations about the growth axis and also substantial rotations about the other two axes. Rotations about the growth axis arise from substrate-induced strain, so choosing substrates with smaller lattice parameter would be desirable. Introduction of rotations about the orthogonal axes may be done by replacing the La with a smaller counterion such as Y. 

Our study has certain limitations. The calculations employ a frontier orbital model which includes only the $t_{2g}$-derived antibonding bands. DFT+DMFT calculations based on correlated atomic-like $d$-states embedded in the manifold of non-correlated oxygen states provide a more fundamental description. Our previous work\cite{PhysRevB.87.155127} indicates that the two models give very similar results if both calculations are tuned so that bulk LaVO$_3$ is a Mott insulator, but the implications of the full (but computationally very heavy) DFT+DMFT procedure for the superlattice problem remain an open problem for future research. Further, our calculations are based on the single-site DMFT approximation, which includes all local effects but misses inter-site correlations. While it is generally accepted that these calculations give the correct trends and qualitative behavior, the quantitative accuracy of the methods is not known.  Unfortunately, as yet cluster extensions of DMFT are prohibitively expensive for the multiband models considered here.

The experimental results of L\"uders et. al.\cite{PhysRevB.80.241102} therefore provide an interesting challenge to materials theory. They indicate that superlattices display ferromagnetism when the corresponding bulk solid solutions do not, whereas the present state of the art of real materials dynamical mean field calculations suggests that superlattices should be less likely to display magnetism than the corresponding bulk solid solutions. This discrepancy requires further investigation.

\section*{Acknowledgements}
We thank U. L\"uders and J. Okamoto for helpful conversations. We acknowledge support from DOE-ER046169. HTD acknowledges partial support from Vietnam Education Foundation (VEF). We acknowledge travel support from the Columbia-Sorbonne-Science-Po Ecole Polytechnique Alliance Program and thank Ecole Polytechnique (HTD and AJM) and J\"ulich Forschungszentrum (HTD) for hospitality while portions of this work were conducted. A portion of this research was conducted at the Center for Nanophase Materials Sciences, which is sponsored at Oak Ridge National Laboratory by the Scientific User Facilities Division, Office of Basic Energy Sciences, U.S. Department of Energy. We use the code for CT-HYB solver\cite{PhysRevLett.97.076405} written by P. Werner and E. Gull, based on the  ALPS library.\cite{Albuquerque20071187}

\appendix*
\section{Lattice constant and V-O bond length ratio}

\begin{table*}
\begin{center}
\caption{\label{table:P21m_struct} Wyckoff positions and lattice constants for LaVO$_3$ with $P2_1/m$ structure from our calculation based on tilt angles taken from the experiment (Ref.~\onlinecite{PhysRevB.85.184101}). La positions are fixed manually but do not affect lattice constants or $c/a$ ratio. In our notation, $d$ is the V-O bond length, $a_0,b_0$ and $c_0$ are lattice constants, $\beta_0$ is the angle between $a_0$ and $c_0$, $a$ and $c$ are pseudocubic lattice constants (growth direction is along $c$ direction, $a$ is perpendicular to $c$).}
 \begin{ruledtabular}
\begin{tabular}{cccccccc}
Atom        & $x$   &  $y$   & $z$    & Atom        & $x$   &  $y$   & $z$     \\
\hline
La$^{(1)}$  & 0     &  0.25  & 0      & La$^{(2)}$  & 0.5   &  0.25  &  0.5    \\
V$^{(1)}$   & 0.5   &  0     & 0      & V$^{(2)}$   & 0     &  0     &  0.5    \\
O$_1^{(1)}$ & 0.4662&  0.25  & 0.0660 & O$_1^{(2)}$ & 0.0392&  0.25  &  0.4392 \\
O$_2^{(1)}$ & 0.7638& -0.0138& 0.2362 & O$_2^{(2)}$ & 0.2652& -0.0493&  0.2652 \\ 
\\
            &$a_0$(\AA)&$b_0$(\AA)& $c_0$(\AA)& $\beta_0$  & $a$ (\AA) & $c$ (\AA) & $c/a$ ratio \\
exp. LVO thin film\cite{PhysRevB.85.184101} & 5.55 & 7.82 & 5.55 & $89.489^\circ$ & 3.91 & 3.945 & 1.008 \\
exp. superlattice\cite{PhysRevB.80.241102,PhysRevB.83.125403} & NA & NA & NA & NA & 3.88 & 3.95 & 1.018\\
calculated with $d=2$\AA  & 5.5988      & 7.8290   & 5.5821 & $88.9732^\circ$ & 3.915 & 3.988 & 1.019 \\
calculated with $d=1.983$\AA  & 5.5512      & 7.7623   & 5.5346 & $88.9732^\circ$ & 3.881 & 3.954 & 1.019 \\
 \end{tabular}
  \end{ruledtabular}
 \end{center}
\end{table*}

In this appendix, we present a more complete discussion of the strain-induced lattice distortions. The in-plane lattice constant of a superlattice epitaxially grown on a substrate matches that of the substrate and may therefore be different from the lattice constant preferred in a free-standing film or bulk material. The out-of-plane lattice constant is typically free to relax, and in the presence of an in-plane strain may also be different from that found in bulk materials. 

A difference in V-V distance may arise from a change in V-O bond length or from a difference in buckling of V-O bonds.  We consider both possibilities here, but first remark that  the main differences in structure between bulk and  experimentally studied superlattices arise from differences in octahedral rotation. In the experimentally-studied superlattices, the in-plane V-V distance  is in fact slightly less than the V-V distance in LVO. The V-O bond lengths have not been measured for the superlattice, but to a high degree of accuracy  we are able to reconstruct the measured superlattice using the measured tilt angles given from experiments\cite{PhysRevB.85.184101} structure, assuming that all  V-O bond lengths are equal.  Assuming the $P2_1/m$ structure, we varied the in-plane and out-of-plane V-O bond lengths to fit the experimental data and found that $c/a\approx1.02$ only when the mean bond length $d$ is found in the range from  $1.983$ to $2$\AA~depending on which experimental result is fit but in all cases the V-O bond lengths are found to be equal to within an accuracy of $0.3\%$. Therefore, we believe that all the V-O bond lengths should, to a good approximation, be the same.  The structure used in our calculations is presented in Table.~\ref{table:P21m_struct}. Although there are slight mismatches in in-plane angle and lattice constants, the $c/a\approx1.02$ ratio and bond angles are compatible with the experiment.

\begin{figure}[h]
 \includegraphics[width=\columnwidth]{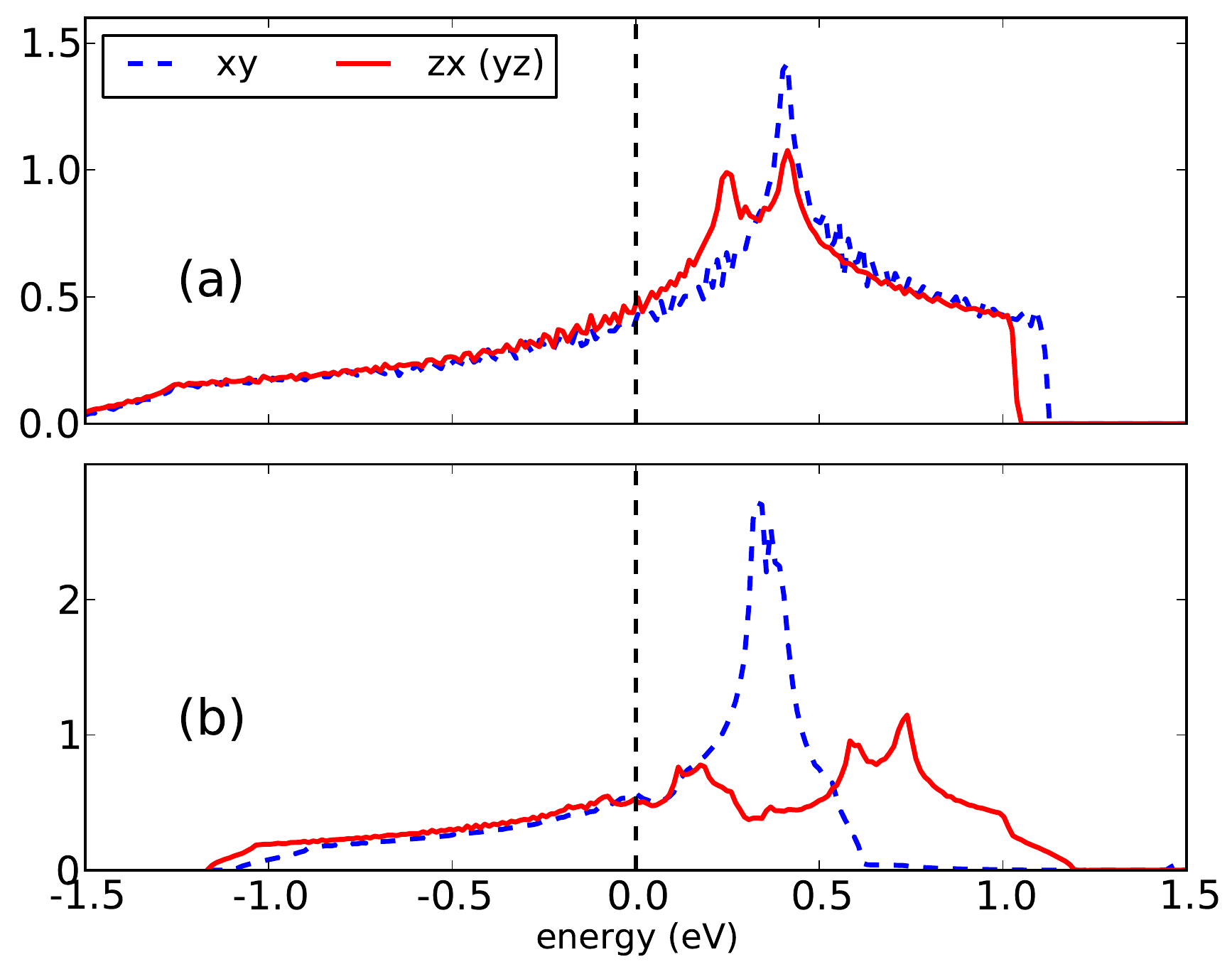}
 \caption{\label{fig:bondlength_pDOS}(Color online) Partial DOS for bulk LVO with $c/a=1.02$ with the $c/a$ ratio due to a change in V-O bonds  (panel (a)) and to $P2_1/m$ lattice structure with $\alpha=\beta=0,\gamma=11.5^\circ$ (similar to superlattice structure)(panel (b)). The dashed blue curve is the $xy$ orbital, the solid red curve is the degenerate $yz$ (or $zx$) orbital. The dashed vertical line marks the Fermi level.}
\end{figure} 

Changing the amount of rotation has a  different effect on the electronic structure than does changing the ratio of V-O bond lengths. Fig.~\ref{fig:bondlength_pDOS} compares the partial DOS for the two cases, using as example  a hypothetical  LaVO$_3$ crystal  with $c/a=1.02$. The upper panel presents the DOS for the untilted structure with straight V-O-V bonds and the $c/a$ ratio induced by a difference in in-plane and out-of-plane V-O bond lengths. The lower panel presents the case of all equal V-O bonds, with the $c/a$ ratio produced by octahedral rotations about the $z$ axis. The densities of states are quite different, but can be understood from  the simple energy dispersion
\begin{equation}
\begin{aligned}
\epsilon_{xy}(\mathbf{k}) &= 2t_\parallel(2-\cos k_x-\cos k_y) + \\
      &+4t_\parallel'(1-\cos k_x\cos k_y),\\
\epsilon_{xz}(\mathbf{k}) &= 2t_\parallel(1-\cos k_x) + \\
      &+2t_\perp(1-\cos k_z) + 4t_\perp'(1-\cos k_x\cos k_z),
\end{aligned}
\label{eqn:dispersion}
\end{equation}
where $t_\parallel$ and $t_\perp$ are the in-plane and out-of-plane nearest neighbor hopping integrals and $t_{\parallel,\perp}'$ are the second neighbor hoppings.  The lower band edge is assumed to be the same for all orbitals but we assume that  the lattice distortions lead to different values for the in-plane and out of plane hoppings.

\begin{figure}[h]
 \includegraphics[width=\columnwidth]{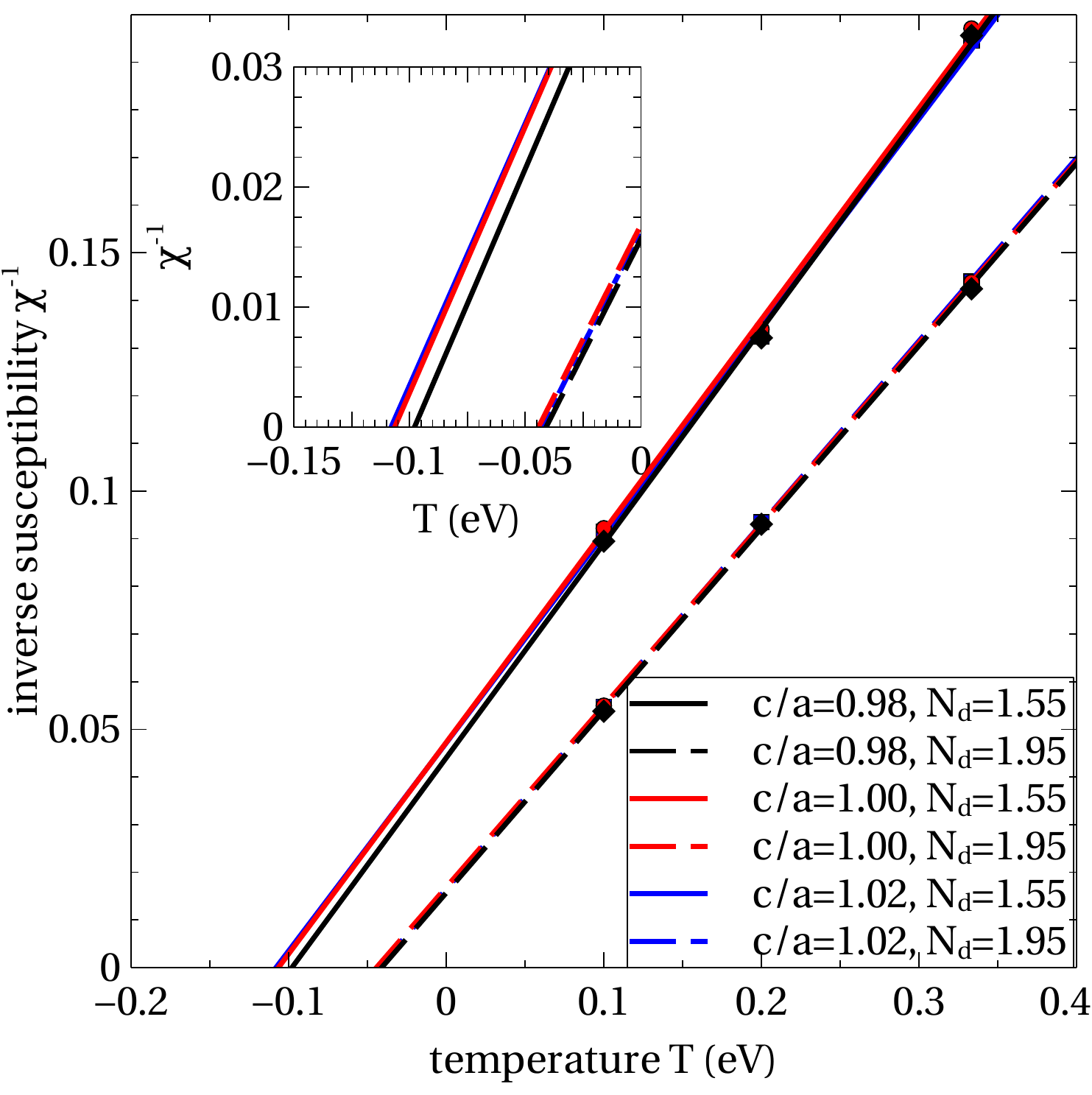}
 \caption{\label{fig:chi_bondlength}(Color online) Inverse susceptibility vs. temperature for cubic structure of bulk hole-doped LVO. The in-plane and out-of-plane bondlengths are changed so that the octahedral volume is unchanged: tensile strain ($c/a=0.98$ - black lines), no strain ($c/a=1.00$ - red lines) and compressive strain ($c/a=1.02$ - blue lines). Two levels of hole doping are considered: $n=1.55$ (solid lines) and $n=1.95$ (dashed lines). These lines are linear fits for the data points.}
\end{figure} 

The lower band edge is defined to be zero and is independent of the distortion. The energy of the upper edge of the $xy$ band is $\epsilon_{xy}(\pi,\pi) = 8t_\parallel$ and of the $xz/yz$ bands is  $\epsilon_{xz}(\pi,\pi) = 4t_\parallel + 4t_\perp$. The positions of the van Hove singularities are at $\mathbf{k}=(0,\pi)$ or $(\pi,0)$. For $xy$ band there is only one van Hove peak,  at $\epsilon^V=4t_\parallel+8t'$; while for $xz$ band, there are two van Hove peaks at $\epsilon_1^V=\epsilon^V$ and $\epsilon_2^V=4t_\perp+8t'$. When $t_\perp$ is different from $t_\parallel$, the difference in bandwidth of $xy$ and $zx$ orbitals is $4t_\parallel-4t_\perp$, which is also the distance between the two van Hove peaks of $zx$ band $\epsilon_1^V-\epsilon_2^V$. 

With these definitions, we are in a position to understand the changes in the band structure. When the V-O bond lengths change (Fig.~\ref{fig:bondlength_pDOS}a) so that the $z$-bond is longer and the in-plane bond is shorter but the octahedral volume is unchanged, the band structure calculation indicates that $t_\perp$ decreases but $t_\parallel$ increases slightly. The difference between the bandwidth of the $xy$ and $xz$ bandwidths is $4t_\parallel-4t_\perp$ which is the same as the splitting between the van Hove peaks in the $xz/yz$ bands. On the other hand, if the $c/a$ ratio is produced by rotation,  (Fig.~\ref{fig:bondlength_pDOS}b), the change is opposite. The in-plane hopping $t_\parallel$ decreases because of the buckled in-plane V-O-V bonds, while the out-of-plane hopping $t_\perp$ is unchanged. The $xy$ band therefore narrows substantially relative to the $xz/yz$ bands. In addition the splitting of the van Hove peaks is greater.  From the bandwidth of $xy$ and $zx$ bands (Fig.~\ref{fig:bondlength_pDOS}b), $t_\parallel\approx0.225$eV, $t_\perp\approx0.35$eV, the van Hove peak distance is $\approx0.5$eV, which is compatible with the peak positions shown in Fig.~\ref{fig:bondlength_pDOS}b.

We tested with DMFT calculations for the Curie temperatures with the V-O bondlength changed. Fig.~\ref{fig:chi_bondlength} is the temperature-dependent inverse susceptibility derived from DMFT for the bulk cubic structure with the $c/a$ ratio changing from 0.98 (tensile strain) to 1.02 (compressive strain). For all the levels of hole doping under consideration, the results are nearly the same for every case of $c/a$ ratio. We conclude that even when the V-O bondlength changes within the physical range, the ferromagnetism is not affected. However, we also found that when the V-O bondlength is such that $c/a\ge1.06$ or $\le0.90$, there is large orbital polarization and the ferromagnetism can be largely affected. But that range is unphysical and can be neglected in the context of this work.

\end{document}